\documentstyle[preprint,aps]{revtex}

\begin{document}
\draft

\title{\bf
On the IMF Multiplicity in $Au+Au$ Reactions}

\author{
Akira Iwamoto$^{1}$, Koji Niita$^{1,2}$, Toshiki Maruyama$^{1}$ 
and Tomoyuki Maruyama$^{1}$ \\[1ex]
$^1$ Advanced Science Research Center, \\
Japan Atomic Energy Research Institute, \\
Tokai, Ibaraki, 319-11 Japan \\[1ex]
$^2$ Research Organization for Information Science and Technology, \\
Tokai, Ibaraki, 319-11 Japan  }

\maketitle

\begin{abstract}
     Intermediate mass fragment (IMF) multiplicity has been investigated for 
$Au+Au$ reactions at incident energies of $100, 250$ and $400 MeV/A$.   From the 
analysis of the impact-parameter-dependence of the IMF multiplicity using our 
QMD plus statistical evaporation model, we found that  1)  statistical decay 
process modifies the results greatly, and 2) the Fermi motion plays a role to 
increase the IMF multiplicity for whole impact-parameter range.
\end{abstract}


\newpage

Intermediate mass fragment ( IMF ) emission is one of the central interest in 
heavy-ion reaction study because the observation of IMF is a good tool to study 
the dynamics and statistics of the violent heavy-ion collisions.  There are many 
experimental data indicating a "multi-fragmentation"  nature of the phenomenon 
but the total understanding of it is not yet achieved.  What is clear is that we 
need a mechanism which is not described by the standard two-body reaction  plus 
statistical evaporation mechanism.   Not only the reaction mechanism but 
also the equation of state of the nuclear matter might be closely related to the 
IMF emission phenomenon.

Among various theoretical models, the molecular dynamics approach \cite{ai91} 
occupies a special position with respect to its many-body dynamical nature.  The 
property and performance of the model is not so clear, however, when both target 
and projectile are very heavy.  There are not many models published so far for 
these very heavy systems because  the calculation is so time-consuming and in 
addition, the stability of the ground state is hard to be achieved.  In this 
report, we will attach the IMF multiplicity for very heavy system using QMD plus 
statistical decay model \cite{ni95,ch95} which has several differences from the 
forgoing models. 

Our concern is to check the ability of the molecular dynamics model to 
understand the multifragmentation reaction at relatively low incident energies. 
 For that purpose, we adopted the Au+Au collisions for the incident energies per 
nucleon of $100, 250$ and $400 MeV$.  For these reactions, there are published 
data of intermediate-mass-fragment (IMF, defined as fragment of ${3 \leq Z \leq 
30}$ ) multiplicities as a function of the impact parameter \cite{ts93}.  These 
data are suitable for the check of the reaction model because the impact 
parameter dependence with three different incident energies gives detailed 
information to the reaction mechanism.   In \cite{ts93}, these data were 
analyzed with QMD \cite{pe89} and QPD \cite{bo88} models, both of them based on 
the molecular dynamics approach.  It was found that these two models 
underestimated the data much, especially at large impact parameters.  To get rid 
of this failure, they tried \cite{ts93} to connect two statistical cluster 
emission models.   One is the combination of QMD \cite{pe89} and statistical 
multifragmentation model \cite{bo85} and the other is QPD \cite{bo88} and 
expanding evaporating source model \cite{fr90}.  These combined models are able 
to yield increased multiplicity at large impact parameters.  However, the impact 
parameter dependence for $100MeV$ incident energy was totally broken with the 
addition of these cluster emission models.  From these observations, the 
conclusion \cite{ts93} was that the ability of the molecular dynamics approach 
to understand the multifragmentation was questionable.

We think, however, that we should try more extended survey of the performance of 
the molecular dynamics model before coming to the final conclusion.  For that 
purpose, we will present calculations based on our model of molecular dynamics.  
We use exactly the same model as was discussed in \cite{ni95,ch95}.  This model 
proved to reproduce the experimental data well for nucleon-induced reactions 
from about $100 MeV$ up to $3 GeV$ incident energies.   By combining the QMD 
model and the standard statistical model, they were able to reproduce the 
emission of particles from $1MeV$ to several $GeV$.  There are essentially two 
points in this model which differ from the foregoing models used in \cite{ts93}. 
  One is that this model produced the ground state of the heavy nuclei without 
introducing the Pauli potential.  Therefore, all the nucleons are moving around 
which is in contrast with the models of \cite{pe89,bo88} where the internal 
velocities of the nucleons are very small due to the way of constructing the 
ground state.  The second point is that we combine the statistical model 
\cite{ni95}, which is based on the simple evaporation model, to our molecular 
dynamics calculations.  We don't assume any fragmentation mechanism in this part 
of the model because our concern is to check the ability of the dynamical model 
to describe the multifragmentation.  Another reason not to introduce the 
statistical fragmentation model is because it's hard to avoid the double 
counting between dynamical part and statistical part.  It will turn out that the 
influence of the standard statistical model we used in this paper is very large 
and it changes the results greatly, which will be one of our main conclusions of 
this paper.

The details of our model are given in \cite{ni95,ch95} and we will not repeat 
them here but point out some essences:  A nucleon is represented by a Gaussian 
wave packet of the width fixed as ${L=2.0fm^2 }$ (cf.Eq.(1) of \cite{ni95} ) 
and 
the total wave function is assumed as a direct product of these single-particle 
wave functions.  The basic equations of motion are determined from the 
time-dependent variational principle where the basic Hamiltonian is composed of 
the Skyrme and Coulomb interactions and symmetry potential.   The soft EOM which 
gives compressibility of $237.3MeV$ is adopted but the effect caused by 
switching to hard EOM will also be discussed in this paper.  The coefficient of 
the symmetry potential is fixed as $Cs=25MeV$ (cf. Eq.(5) of \cite{ni95} ).  No 
momentum-dependent interaction is included and the ground state is  prepared 
with the condition that the experimental binding energy is reproduced.  We 
should mention here that if we include the momentum dependent term, it's almost 
impossible to prepare the ground state of very heavy nuclei without introducing 
the Pauli potential or other stabilizing mechanisms.  For the collision term, 
included are the elastic scattering of nucleons and pions, and inelastic 
channels of $\Delta $ and $N^* $ resonances are treated explicitly (cf.Eq.(9) 
of \cite{ni95} ).   For the other details of the inelastic collisions, see 
\cite{ni95}.  The relativistic kinematics are adopted and in addition, the 
argument of the interaction  is defined as Lorentz scalar quantity which has a 
merit of compensating the unreasonably large Lorentz contraction effect(cf. 
Eqs.(29-42) of \cite{ni95}).

Calculations are done by counting the number of IMF (${3 \leq Z \leq 30}$).  For 
the classification of fragments, we used a minimum distance chain procedure, 
i.e., two nucleons are considered to be bound in a fragment if the distance 
between their centers of Gaussian wave packets is smaller than $4.0 fm$.  We 
checked the dependence of the final results on this minimum distance and found 
that it causes little change if the switching time from the QMD calculation to 
statistical model calculation is not too short.  The standard switching time 
used in \cite{ni95} for nucleon induced reactions was $100fm/c$.  In the present 
case of $Au+Au$ system, this choice is a little too short and we fixed the time 
as 
$200fm/c$, which value was also used in \cite{pe89}.   In $100fm/c$, the 
fragment separation for low impact parameters is not finished yet, especially 
for low incident energies.  The choice of switching time $100fm/c$ therefore 
causes the change of the final results depending on the minimum distance.  For 
the choice of the  switching time between $200fm/c$ to $400fm/c$, we found that 
the physical quantities calculated in this paper is rather stable, the deviation 
is within a few percent.  The same statistical model as \cite{ni95} was used 
after this switching time in which the emissions of ${n, p, d, t, ^{3}He}$ and 
$^{4}He$ are explicitly included.  The inclusion of heavier fragments is 
expected to  modify the result somewhat but not much.

In fig.1, we plot the mean IMF multiplicity as a function of the impact 
parameter b for three different incident energies.  Experimentally, the impact 
parameter b is determined with the total charge multiplicity.  IMF multiplicity 
data are plotted with solid circles with error bars \cite{ts93}.   Our  
calculation at $t=200fm/c$ using the QMD output is shown by  solid lines.  For 
comparison, two other theoretical calculations performed in \cite{ts93} are 
given: one is the QMD model calculation \cite{pe89} which is shown by the dashed 
lines and the other is the QPD model calculation \cite{bo88} which is shown by 
the dash-dotted lines, both are taken from Fig.2 of \cite{ts93}.  The filtering 
of the experimental acceptance is not included in all calculations given in this 
paper, since the filter was not available for us.  The change caused by the 
filter is seen in figs.2 and 3 of \cite{ts93} for four kinds of calculations 
given there.  One has to keep in mind of this fact in the following discussion 
on the comparison between calculations and data.

For $E/A=100MeV$, we achieved a good reproduction of the data, which is much 
better than other two theoretical calculations.  For two higher energies, our 
calculations deviate from the data in two ways: one is the shift of the peak to 
lower energy and another is the overestimation of IMF at low impact parameters.  
The shift of peak to lower energies seems to be common to other two theoretical 
calculations.  It looks that the shift of our calculation is not so large as 
compared to other two calculations.  On the other hand, the overestimation of 
the multiplicity at low impact parameters are not seen in these two 
calculations.  They underestimate the data for whole impact parameter values.  
As a whole, our calculations gives a larger IMF multiplicity than other two 
model calculations.

A rather large deviation of our results compared with other two model 
calculations is the point of interest.   Out results are nearer to the QMD of 
\cite{pe89} but the IMF multiplicity is larger in our case, especially at low  
and high impact parameter values.  To our regret, the calculation of  
$E/A=250MeV$ is lacking for the model of \cite{pe89} in \cite{ts93} and the 
systematic comparison is not perfect.  When we 
compare our results with those of QPD \cite{bo88}, we observe still larger 
difference.  The reason of these differences is not very clear but we think the 
following consideration is important.  As we mentioned earlier, the main 
difference of our model to other two models is that they included Pauli 
potential.  As a result, the velocity of each nucleon has zero or almost zero 
value at the ground state although it has finite momentum value.  The Fermi 
motion is much suppressed in their treatment compared with ours and thus no 
wonder that 
fragmentation dynamics also changes.   We cannot say which model is better at 
this stage but it looks that Fermi motion has tendency to increase the IMF yield 
in these reactions.  The effect of Fermi motion was clearly observed in the 
nucleon-induced reaction \cite{ch95}.

In fig.2, we show our results of the b-dependence of IMF multiplicity after we 
transferred the results of our QMD output at $200fm/c$ to statistical 
evaporation code and calculated the decay chains, which is shown by the solid 
lines.  We gave the same data as in fig.1 and also two other model calculations 
taken from fig.3 of \cite{ts93}: one is QMD \cite{pe89} plus statistical 
multifragmentation model (SMM) \cite{bo85} which is shown by the dashed lines 
and the other is QPD \cite{bo88} plus expanding evaporating source (EES) model 
\cite{fr90} which is shown by the dash-dotted lines.  

First of all, we observe a very big change caused by our statistical evaporation 
model by comparing with fig.1.  Our statistical  model is a standard evaporation 
model which assumes no special fragmentation mechanism.  Therefore the change of 
our calculations from fig.1 to fig.2 means that the decay of excited fragment by 
evaporating light charged particles and neutron is quite important to count the 
IMF multiplicity.  The raw QMD results should not be compared with experimental 
data.  As a result of this evaporation, our calculated values are reduced quite 
much, especially for the low impact parameters.  The overestimation at low 
impact parameters and a resultant skewed b-dependence shape observed in fig.1 no 
more exist.  The shape of the b-dependence becomes very good for all three 
incident energies.  The problem is the overall underestimation of the data and a 
shift of the b-dependence shape to lower b values.  In contrast, other model 
calculations show a very different behavior.  For $E/A=400MeV$, many IMF are 
produced in the statistical calculation stage and they overestimate the data, 
especially in QMD plus SMM model calculations.  One should notice, however, 
that this 
model reproduces the data well after the filter is applied as is seen in Fig.3 
of \cite{ts93}.  The QPD plus ESS model underestimates the data at low and high 
b values.  At $E/A=100MeV$,  QMD plus SMM model gives larger deviation from data 
than QMD alone which is shown in fig.1.  It is hard to draw conclusion for these 
other 
calculations but we think our results give quite a reasonable b-dependence 
behavior and therefore, our OMD plus statistical decay model hits some basic 
mechanism of IMF production.  We should find a reasonable explanation, 
however, why the IMF value is about the half of the experimental data.  In this 
respect, we tried the calculations with different Skyrme interaction parameters 
which give a hard equation of state with the compressibility of 380MeV.  The IMF 
multiplicity obtained with this interaction, however,  changed only a little 
from the one obtained with the soft equation of state.  The typical change was a 
few percent and thus the equation of state is not the origin of the 
underestimation of IMF multiplicity.   

In concluding, from the analysis of the impact parameter dependence of IMF 
multiplicity  for $Au+Au$ collisions at $ E/A=100,250 $ and $ 400 MeV $ with QMD 
plus statistical evaporation model, two important findings are obtained.  First 
is the necessity of calculating the statistical decay of the excited fragments 
produced in the molecular dynamics calculations before comparing with the 
experimental 
data.   Inclusion of this greatly changes the results, both for the b-dependence 
and for the absolute values of IMF.   Final b-dependent shape resembles the data 
but absolute values are about the half of the data.  Second is the importance of 
the Fermi motion for whole region of impact parameters.  Suppression of Fermi 
motion caused by introducing Pauli potential has tendency to result in a further 
shortage of calculated IMF multiplicities.

\newpage
\noindent
{\Large Figure Captions}
\bigskip
\begin{itemize}

\item[Fig.1]
IMF multiplicities for three incident energies of $Au+Au$ reactions plotted as 
functions of the impact parameter b.   Data \cite{ts93} are shown by the solid 
points and our QMD calculations are shown by the solid lines.   Dashed lines and 
dash-dotted lines depict the QMD \cite{pe89} and QPD \cite{bo88} calculations, 
respectively, both taken from \cite{ts93}.

\item[Fig.2]
Same as fig.1 for data points.  Solid lines represent our IMF multiplicities 
after statistical decay.  Dashed lines and dash-dotted lines depict the QMD+SMM 
and QPD+EES calculations, respectively, taken from \cite{ts93}.

\end{itemize}

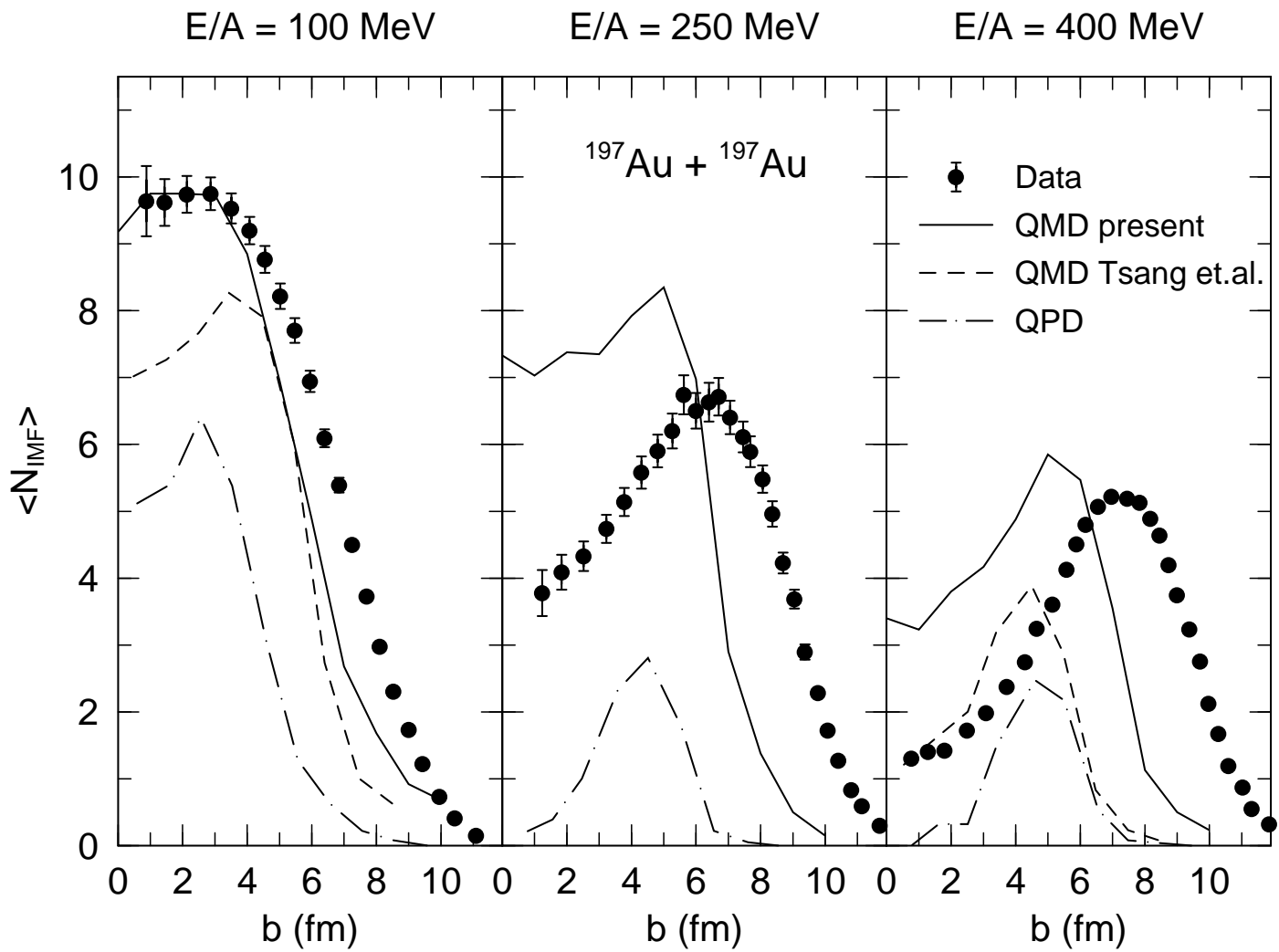
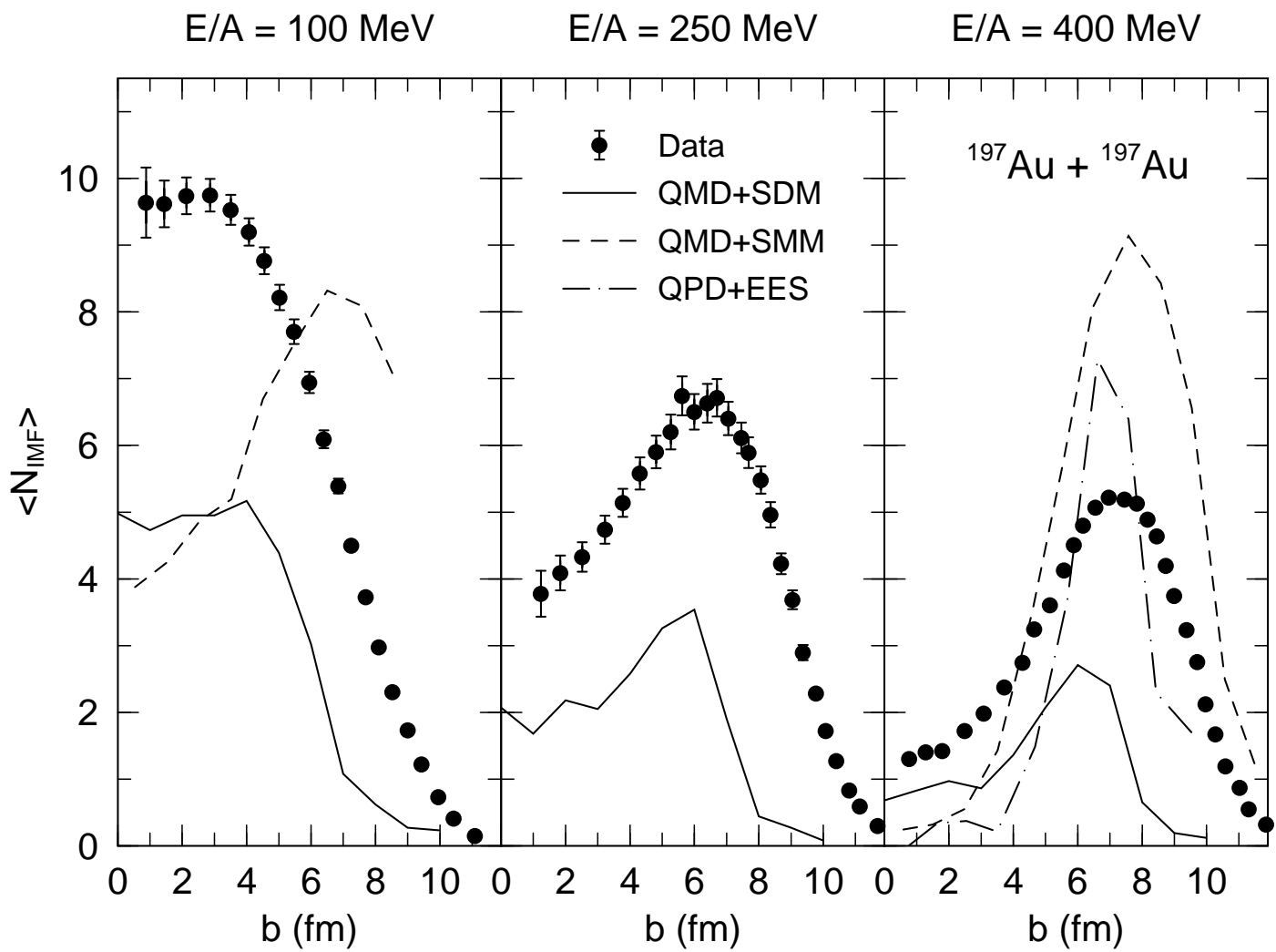
\end{document}